\documentstyle[aps,multicol,epsf,prb]{revtex}
\begin{document}
\draft
\title{Giant Shapiro steps for two-dimensional Josephson-junction
  arrays with time-dependent Ginzburg-Landau dynamics}
\author {Beom Jun Kim and  Petter Minnhagen}
\address {Department of Theoretical Physics,
Ume{\aa} University,
901 87 Ume{\aa}, Sweden}
\preprint{\today}
\maketitle
\thispagestyle{empty}
\begin{abstract}

Two-dimensional Josephson junction arrays at zero temperature are 
investigated numerically within the resistively shunted junction (RSJ)
model and the time-dependent Ginzburg-Landau (TDGL) model with global
conservation of current implemented through the fluctuating twist boundary 
condition (FTBC). 
Fractional giant Shapiro steps are found for {\em both} the RSJ and TDGL cases. 
This implies that the local current conservation, on which the RSJ model 
is based, can be relaxed to the TDGL dynamics with only global current
conservation, without changing the sequence of Shapiro steps. However, when the 
maximum widths of the steps are compared for the two models
some qualitative differences are found at higher frequencies. 
The critical current is also calculated
and comparisons with earlier results are made.
It is found that the FTBC is a more adequate boundary condition than 
the conventional uniform current injection method because it minimizes
the influence of the boundary.
\end{abstract}


\begin{multicols}{2}
  Two-dimensional (2D) Josephson junction arrays (JJA's)
have been the subject of much current interest 
because of their phase transitions and nonequilibrium transport properties.~\cite{review}
In the presence of an external uniform magnetic field, the 
frustration $f$, defined by the number of flux quanta per plaquette,
plays an important role reflected, e.g., in the value of the critical 
current.~\cite{teitel} Furthermore, when an 
$L \times L$ square array with $f=p/q$ ($p,q$ are integers)
is driven by combined direct and alternating currents
$I_{\rm ext}(t) = I_d + I_a\sin \omega t$,
fractional giant Shapiro steps at voltages 
\begin{equation}
\langle V \rangle = n \left( \frac{\hbar \omega}{2e}\right)\left(\frac{L}{q}\right), 
\end{equation}
where $\langle \cdots \rangle$ is the time average and $n$ is an integer, have 
been observed both in experiments~\cite{benz-l} and in computer 
simulations.~\cite{khlee,free,octavio,tiesinga1}
Qualitative arguments have been proposed 
to explain these fractional steps in terms of vortex 
motion~\cite{khlee,free} and topological invariance.~\cite{kvale,mychoi}
In addition to the fractional steps, a series of small subharmonic steps 
has been found.~\cite{skim}

Two slightly  different models have been used to catch the
essential properties of a JJA: the resistively shunted junction (RSJ)
model and the time-dependent Ginzburg-Landau (TDGL) model.~\cite{f1} 
The RSJ
model is based on the assumption that all the current goes through the
array and that the current is conserved locally at each instant. The
TDGL model in the absence of an external current describes either a situation where all the current goes
through the array, but where the local current conservation is relaxed,
or a situation where not all the current goes
through the array (leakage to the ground) and the current is conserved
at each instant. The former view means that the TDGL model can be
regarded as a simplified version of the RSJ model and at the same time
as a less restrictive model of a JJA.
The latter view has led to the suggestion that a JJA with {\it} local
damping is a possible realization of the 
TDGL model.~\cite{beck,tiesinga} In the presence of an external current
the physics of the TDGL model depends on the choice of the boundary
condition. We use here a boundary condition corresponding to
the case when the normal current flow is through the array just
as in the RSJ case. However, current is only
conserved globally and not locally. In this way, we can compare the effects
caused by the difference in local current conserved dynamics
as in the RSJ case with the TDGL dynamics which only has
global current conservation.~\cite{note1}

Both dynamic models are equivalent as far as static
equilibrium properties are concerned,
since they have the same equilibrium Boltzmann distribution.~\cite{bmo}
On the other hand, for dynamic quantities such
as the dynamic dielectric function,~\cite{bmo,jonsson}
flux-noise spectrum,~\cite{tiesinga,hwang} and current-voltage ($I$-$V$)
characteristics,~\cite{bmo,simkin} the equivalence is not
guaranteed. 
It has recently been suggested that the TDGL model could describe the flux-noise
experiment for a JJA better than the RSJ model.~\cite{tiesinga}
However, a somewhat different conclusion was reached in
Ref.~\onlinecite{bmo} where properties like the linear response and
nonlinear $I$-$V$ characteristics were found to be the same for the two
models.

In Ref.~\onlinecite{bmo} a novel boundary condition
[the fluctuating twist boundary condition (FTBC)] based on 
global current conservation was introduced.~\cite{f2}
We show in this work that the very same Shapiro steps
are found in the TDGL and RSJ models when we employ the
FTBC as the boundary condition.~\cite{note1} This suggests
that the existence of the steps does not depend on the details of the dynamic models: 
This robustness can be explained by the topological nature of steps
where the ground state degeneracy (both 
models are equivalent in this respect) has been shown to play an 
important role.~\cite{mychoi}
The widths of Shapiro steps as a function of $I_a$ and $\omega$
have also been a subject of
much interest. For example, the maximum width of the integer and the fractional
steps have been shown to have a different frequency dependence for
larger frequencies.~\cite{octavio,rzchowski,marino} We find that the
maximum width of the half-integer
step for the TDGL dynamics has a different frequency dependence than for
the RSJ case. This offers an experimental possibility (similar to the
experiment on an $f=0$ array in 
Ref.~\onlinecite{ravindran})
to investigate whether a JJA could sometimes be better described by
the TDGL dynamics.

In the presence of applied direct currents, the critical currents
$I_c(f)$, beyond which the voltage takes nonzero values, have been measured
in experiments~\cite{benz-b} and simulations.~\cite{khlee,benz-b}
Although theoretical predictions~\cite{teitel,benz-b}
and experiments~\cite{benz-b} for $f=1/2$ give the result $I_c(f=1/2) 
= \sqrt{2} - 1  \approx 0.414$ in units of $I_c(f=0)$, computer simulations
with the conventional method of uniform current injection 
gives the value 
$0.35 \pm 0.01$.~\cite{khlee,free,octavio,tiesinga1,simkin2} In Ref.~\onlinecite{free} it has 
been argued that this discrepancy is due to the boundary condition and that 
the conventional method destroys the translational symmetry of the 
ground state. 
On the other hand, we find in this work that the FTBC gives the value
0.4142(1)  for both the RSJ and TDGL models,
which suggests that the FTBC is a
more adequate boundary condition since it conserves translational symmetry.

We start by introducing the equations of motion for the RSJ and 
TDGL models with the FTBC (see Ref.~\onlinecite{bmo}
for details). 
In the FTBC the twist variable ${\bf \Delta} \equiv (\Delta_x, \Delta_y)$
is introduced and the gauge-invariant phase difference is changed into~\cite{f3}
\begin{equation} \label{eq_phiij}
\phi_{ij} = \theta_i - \theta_j - A_{ij} - {\bf r}_{ij}\cdot{\bf \Delta},
\end{equation}
where $\theta_i$ is the phase of the superconducting order parameter at site 
$i$, ${\bf r}_{ij} \equiv {\bf r}_i - {\bf r}_j$ is a unit vector from site 
$i$ to $j$, and $A_{ij} \equiv (2\pi/\Phi_0)\int_i^j {\bf A}\cdot d{\bf l}$ 
with the magnetic vector potential ${\bf A}$ and the 
flux quantum $\Phi_0$ for Cooper pairs. 

In the RSJ model, the equations of motion for phase variables are determined by
the {\em local} current conservation at each site (see, for example, 
Refs.~\onlinecite{khlee} and \onlinecite{skim}):
\begin{equation} \label{eq_rsj_theta}
\dot\theta_i = -\sum_j G_{ij}{\sum_k}^{'}\sin\phi_{jk},
\end{equation}
where the primed summation is over four nearest neighbors of $j$, $G_{ij}$ is 
the square lattice Green function, and the unit of time is 
$\hbar/2eRI_c$ with shunt resistance $R$ and critical
current $I_c$ of the single junction.
In this work we only consider the array at zero temperature and accordingly
the thermal noise terms are disregarded (see Ref.~\onlinecite{bmo} for 
finite temperatures). 
For the TDGL model the equations of motion are given 
by~\cite{bmo,jonsson}
\begin{equation} \label{eq_tdgl_theta}
\dot{\theta}_i  = -{\sum_j}^{'}\sin\phi_{ij}, 
\end{equation}
where $t$ is in units of $2e/I_c$.

In the FTBC case, the periodicities of the phase variables are preserved in 
both directions, i.e., $\theta_i = \theta_{i + L{\hat {\bf x}}} = 
\theta_{i + L{\hat {\bf y}}}$, and thus the voltage drop in each
direction across the whole array is given by
\begin{equation} \label{eq_V}
V_x = -\frac{\hbar L}{2e} {\dot\Delta}_x , \ \ \
V_y = -\frac{\hbar L}{2e} {\dot\Delta}_y , 
\end{equation}
from the Josephson relation.
From the condition of {\em global} current conservation in each direction, 
we obtain the equations of motion for the twist variables of 
the array driven by an external current $I_{\rm ext}$ (in units of $I_c$) 
in the $x$ direction:~\cite{bmo}
\begin{eqnarray} 
I_{\rm ext} &=& -\frac{d\Delta_x}{dt} + \frac{1}{L^2}\sum_{\langle ij\rangle_x} \sin\phi_{ij} , 
\label{eq_delta_x} \\
0  &=& -\frac{d\Delta_y}{dt} + \frac{1}{L^2}\sum_{\langle ij\rangle_y} \sin\phi_{ij} ,
\label{eq_delta_y}
\end{eqnarray}
where $\sum_{\langle ij\rangle_x}$ ($\sum_{\langle ij\rangle_y}$) 
is the summation over all nearest-neighboring pairs in the $x$($y$) direction.

It should be observed that we are here considering the TDGL model [as
specified by the dynamical equation~(\ref{eq_tdgl_theta})] 
with globally conserved current [as specified by  
Eqs.~(\ref{eq_delta_x}) and (\ref{eq_delta_y})].  
One can also consider the TDGL model [as specified by Eq.~(\ref{eq_tdgl_theta}) ]  
without global current conservation within the
plane [i.e., without Eqs.~(\ref{eq_delta_x}) and (\ref{eq_delta_y})] 
and interpret it as a model with leakage to the ground.~\cite{beck,tiesinga} 
In this case an applied external current 
only leads to dissipation to the ground at the boundaries where the current
is injected and extracted. Consequently there exist no giant Shapiro steps
for the TDGL model without globally conserved current within the plane.~\cite{thesis}

\begin{figure}
\centerline{\epsfxsize=12.0cm \epsfbox{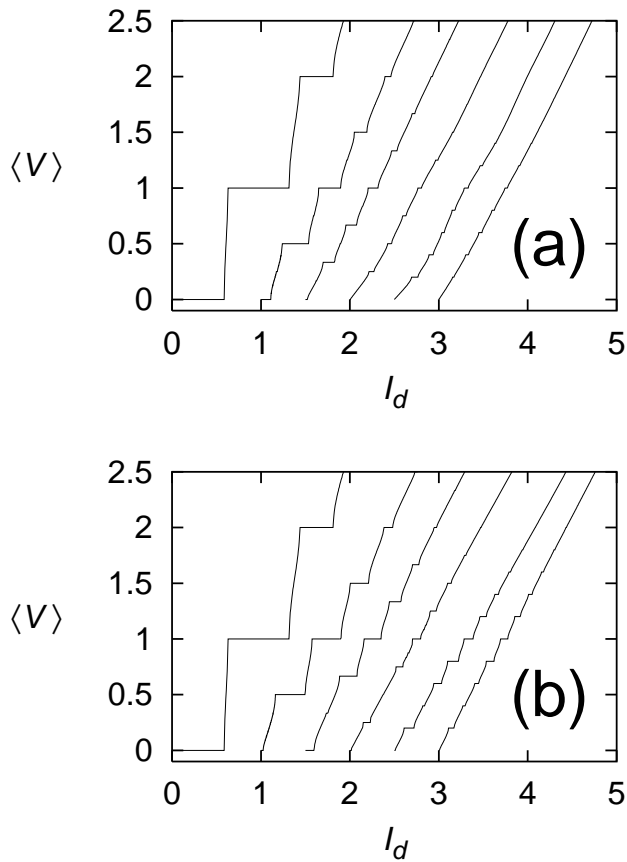}}
\vskip -3.5cm
\caption{Time-averaged voltages $\langle V \rangle$ in units of $L\hbar\omega/2e$
versus direct current $I_d$ for (a) the RSJ and (b) the TDGL models 
in an $L\times L$ Josephson junction array in case of 
$f$ = 0, 1/2, 1/3, 1/4, 1/5, and 2/5 (from the left to the right). 
The sizes of arrays are $L=4$ (for $f=0$),  8 ($f=1/2$ and 1/4), 9 ($f=1/3$),
and 10 ($f=1/5$ and 2/5), and we have used the fluctuating twist
boundary
condition together with the condition of global current conservation 
and applied external currents $I_{\rm ext} = I_d + I_a\sin\omega t$ with $I_a = I_c$ and
$\omega/2\pi = 0.1$ in units of $2eRI_c/\hbar$ for the RSJ model and $I_c/2e$ for the TDGL
model, respectively.
Fractional giant Shapiro steps are clearly shown for the TDGL as well as for the RSJ model. 
All curves except $f=0$ are horizontally displaced for clarity. }
\label{fig_shapiro}
\end{figure}

We first consider an array with an external current 
$I_{\rm ext} = I_d + I_a \sin\omega t$. We use the Euler algorithm with 
discrete time step $\Delta t=0.05$  to integrate the equations of motion 
[Eqs.~(\ref{eq_rsj_theta}), (\ref{eq_delta_x}), and  (\ref{eq_delta_y}) for the RSJ
model, and Eqs.~(\ref{eq_tdgl_theta}), (\ref{eq_delta_x}), and (\ref{eq_delta_y}) for the 
TDGL model], and the time-averaged voltages
$\langle V\rangle$ in units of $L\hbar\omega/2e$ are calculated from
Eq.~(\ref{eq_V}).  We adopt the simulated annealing Monte Carlo method 
to find the ground states of the array and then use them as initial conditions 
of phase variables together with $\Delta_x(t=0) = \Delta_y(0) = 0$.
Figure~\ref{fig_shapiro} shows the fractional giant Shapiro steps in
the $I$-$V$ 
characteristics for (a) the RSJ model and (b) the TDGL model. Although we have found 
quantitative differences in the Shapiro steps for the TDGL model
(e.g., small step sizes at 
$\langle V\rangle =1/3$ and 1/2 for $f=1/3$ and $1/4$, respectively), 
it is clear that with the FTBC not only the RSJ but 
also the TDGL model generates the integer and fractional steps.
We have also observed weak subharmonic steps for both models 
as in Ref.~\onlinecite{skim}.

\begin{figure}
\centerline{\epsfxsize=12.0cm \epsfbox{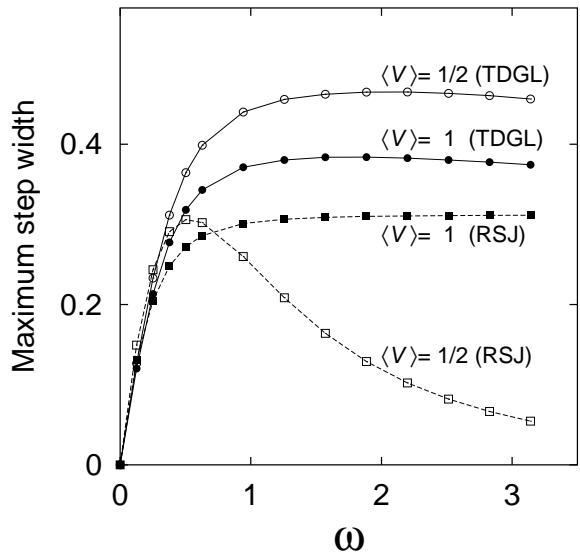}}
\vskip -4.5cm
\caption{Frequency dependence of the maximum widths of the Shapiro steps at
$\langle V\rangle = 1/2$ and $1$ for the RSJ and TDGL models with
the FTBC in case of $f=1/2$.
Our results for the RSJ model are in good agreement with the analytic
results. (Ref.~\protect\onlinecite{rzchowski}).
The high-frequency behavior of the 1/2 step for the 
TDGL model is shown to differ from the RSJ model.
The lines are guides to the eye.
}
\label{fig_width}
\end{figure}

Figure~\ref{fig_width} displays
the maximum widths~\cite{rzchowski,ravindran} of the steps $\langle V\rangle
=1/2$ and 1 versus  $\omega$ for both models. Although both 
show the same qualitative behavior in the low-frequency regime, it is
apparent from the figure  
that the high-frequency behaviors of the half-integer steps are different for 
the two models. Since the frequency dependence of the maximum width
can be measured for a JJA,~\cite{ravindran} this offers the
possibility of experimentally distinguishing between the two types
of dynamics.

We have also performed computer simulations applying a constant direct
current
$I_{\rm ext}(t) = I_d$ for three different cases and obtained the
critical currents $I_c(f)$: One case is the RSJ model with the conventional method of 
uniform current injection which employs
the periodic boundary condition (the free boundary condition)
in the direction perpendicular (parallel) to the applied currents. 
The other two are the RSJ and TDGL models with the FTBC.
We present in Table~\ref{tab} a comparison of these three cases, which reveals
that the FTBC gives correct values 
for both the RSJ and TDGL models.
For the conventional current injection method we obtained different 
values, e.g., $I_c(f=1/2) = 0.35(1)$, as was also found in 
Ref.~\onlinecite{free}. We checked the system size dependence for
$L=4,8, \cdots ,128$ and found no change. Nevertheless, these smaller
values are caused by the boundary condition which destroys the
translational symmetry of the ground state.~\cite{free}
In Ref.~\onlinecite{free} this problem was circumvented 
by a nonuniform injection method which matched the 
translational symmetry of the ground state and the correct  
value $0.414$ was found. In the FTBC case the
translational symmetry of the ground state is automatically preserved
and consequently this boundary condition directly yields the correct result.
We also calculated the critical current with  the busbar 
geometry~\cite{simkin} and obtained an even smaller value of
$I_c(f=1/2)$ than for the conventional uniform injection method, 
as was already noticed in Refs.~\onlinecite{free} and \onlinecite{simkin2}. 
From these comparisons we conclude that the FTBC has an advantage over
other commonly used boundary conditions.

\begin{figure}
\centerline{\epsfxsize=12.0cm \epsfbox{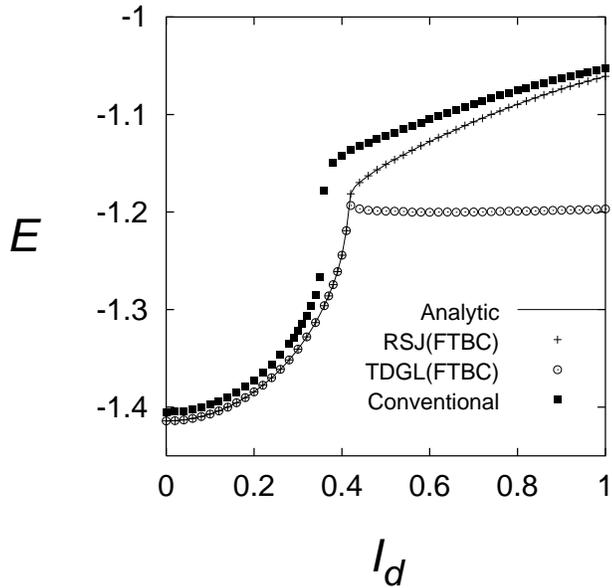}}
\vskip -4.5cm
\caption{Energy per site, 
$E \equiv -\langle\sum_{\langle ij\rangle}\cos\phi_{ij}\rangle/L^2$,  for 
an $L\times L$ array in case of $f=1/2$ with an applied direct current $I_d$.
The results from the analytic equations (solid curve) in 
Ref.~\protect\onlinecite{rzchowski} are in perfect agreement 
with our results for the RSJ model using the FTBC. The TDGL model gives the same $E$ 
below the critical current. However, the RSJ model with the conventional method 
has a cusp structure at a different value of $I_d$. 
}
\label{fig_E}
\end{figure}

\begin{figure}
\centerline{\epsfxsize=12.0cm \epsfbox{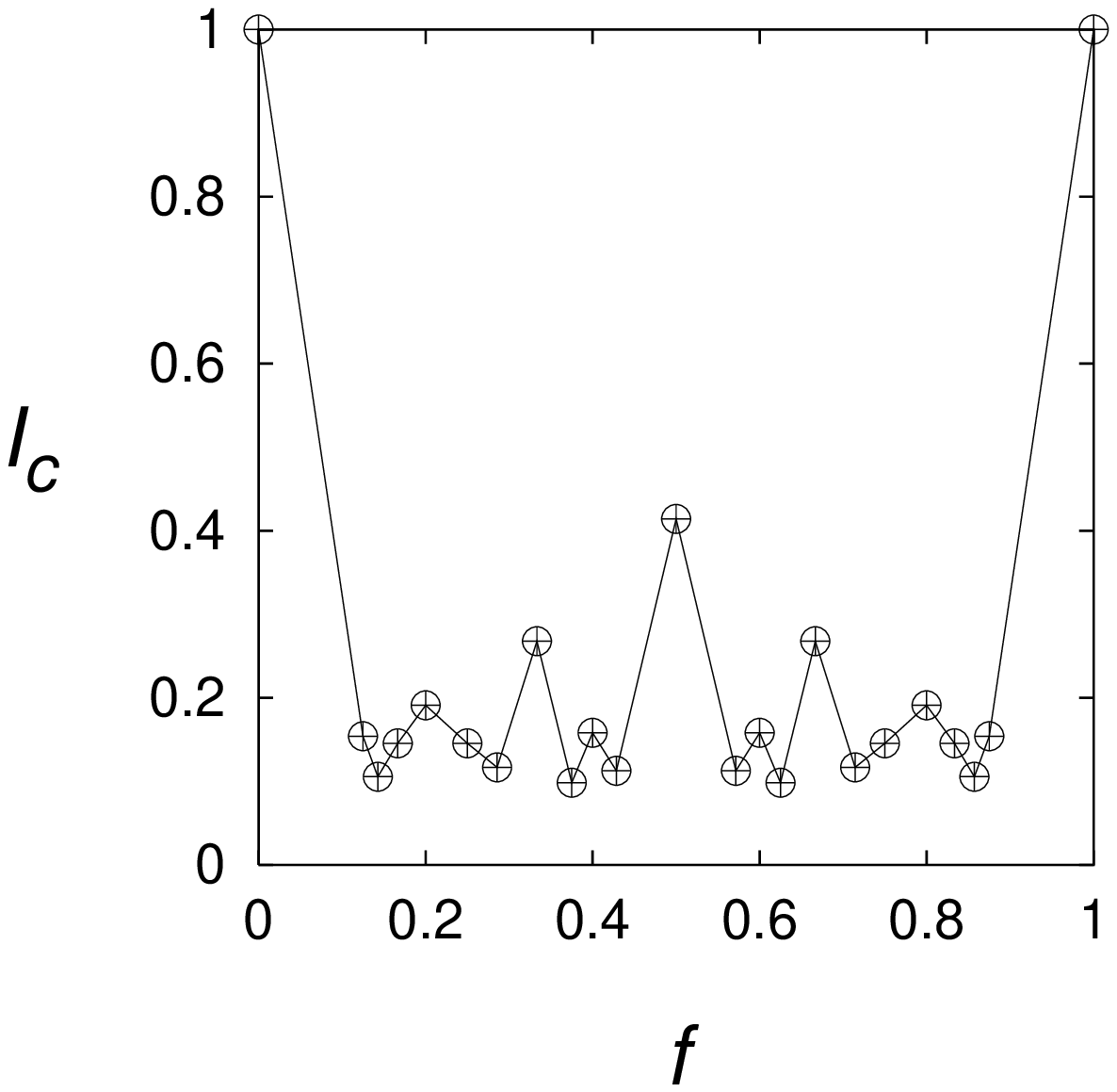}}
\vskip -4.5cm
\caption{Critical currents $I_c(f)$ as a function of $f$ for the RSJ (denoted 
by $+$) and the TDGL ($\circ$) at $f=p/q$ with $q = 1, 2, \cdots, 8$. 
The fluctuating twist boundary condition is used together with the condition of 
global current conservation. For all values of $f$ tested in this work, 
the RSJ and TDGL models give the same value of $I_c(f)$ within numerical 
accuracy.  The line is a guide to the eye.}
\label{fig_ic}
\end{figure}

In Fig.~\ref{fig_E} we show the average energy defined by
$E \equiv -\langle \sum_{\langle ij\rangle}\cos\phi_{ij}(t) \rangle /L^2$ 
as a function of $I_d$ for several cases. 
Our results for the RSJ model with the FTBC are in perfect agreement with the 
results from the analytic equations given in Ref.~\onlinecite{rzchowski}, 
which suggests that our results contain no boundary or
finite-size effects. The TDGL model is found to give 
the same $E$ for currents less than the critical value. 
Beyond the critical current, the TDGL model
gives a lower energy, implying that the array is closer to the
ground state than the RSJ model. 
We believe that this explains the robustness of 
the 1/2 step of the TDGL model at high frequency (see Fig.~\ref{fig_width}),
since it is expected that the ground state and its vortex superlattice 
structure plays an important role in creating the half-integer steps.~\cite{octavio}
We find in all cases that $E(I_d)$ has a cusp 
structure at the critical current.
One may also note in  Fig.~\ref{fig_E} that the conventional uniform
current injection method leads to a result which differs from the
exact analytical result.
Figure~\ref{fig_ic} gives the critical currents $I_c(f)$ at $f=p/q$ with
$q = 1, 2, \cdots ,8$ (for comparisons with previous works, 
see Ref.~\onlinecite{teitel}). For all values of $f$, 
we obtain identical values of $I_c(f)$ for the TDGL and RSJ models.

In conclusion, we have performed simulations for the RSJ and TDGL models 
subject to the FTBC.
Fractional giant Shapiro steps are obtained for {\em both} 
models, which suggests that the existence of the steps does not depend
crucially on the 
condition of instantaneous local current conservation. 
However, the maximum width of the half-integer
step at $f=1/2$ has a qualitatively different high-frequency behavior
for the two models.
The critical currents of the array with direct applied currents were
also calculated for both the models subject to the FTBC and 
compared with the results obtained for the RSJ model 
with the conventional method of uniform current injection.  
It was concluded that the FTBC for both models gives values in agreement with experiments 
and analytic calculations, while the conventional method fails in this respect.

The present calculation supports the conclusion reached in
Ref.~\onlinecite{bmo} that the TDGL  and RSJ models with the FTBC
are qualitatively
equivalent for low frequencies (compare Fig.~\ref{fig_width}) and
small currents (compare Fig.~\ref{fig_E}) whereas for larger
frequencies and larger currents there exist qualitative differences.
The fact that both the models have qualitatively similar sequences of
giant Shapiro steps suggests that the existence of these steps
is strongly linked to an equilibrium property
like the ground state degeneracy.~\cite{mychoi}

B.J.K. wishes to acknowledge
the financial support of the Korea Research Foundation
for the program year 1997. This research was supported by the Swedish
Natural Research Council through Contracts Nos. FU 04040-332 and EG 10376-310.

\narrowtext

\begin{table}

\caption{ Comparison of critical currents at $f=0$, 1/2, and 1/3 for
the RSJ model with the conventional method ($I_c^{\rm conv}$), the RSJ
model with the fluctuating twist boundary condition ($I_c^{\rm RSJ}$), 
the TDGL with the FTBC ($I_c^{\rm TDGL}$), and the analytic results
in Ref.~\protect\onlinecite{benz-b} ($I_c^{\rm anal}$). 
All values are in units of $I_c(f=0)$ and the numbers in parentheses are numerical
errors in the last digits. } 
\label{tab}

\begin{tabular}{c c c c c}
$f$ & $I_c^{\rm conv}$ &$I_c^{\rm RSJ}$ & $I_c^{\rm TDGL}$ &  $I_c^{\rm anal}$ \\ \hline 
0  & 1.00(1)  & 1.0000(1)  & 1.0000(1) &  1          \\
1/2 & 0.35(1) & 0.4142(1)  & 0.4142(1) &  0.41421          \\
1/3 & 0.14(1) & 0.2679(1)  & 0.2679(1) &  0.26789          \\
\end{tabular}
\end{table}

\vskip 2cm
\end{multicols}


\begin{thebibliography}{99}
\bibitem{f3} Similarly to Eq.~(\ref{eq_phiij}), the external current in 
Ref.~\onlinecite{mychoi}
has been included as an effective bond angle term in the gauge-invariant phase difference.
\end{thebibliography}
\end{document}